# Free volume investigation of imidazolium ionic liquids from positron lifetime spectroscopy

Yang Yu*[a,b], Dana Bejan[c], Reinhard Krause-Rehberg[b]

[a]*School of Physics and Optoelectronic Engineering, Nanjing University of Information Science & Technology, Nanjing 210044, Jiangsu, China*
[b]*Institut für Physik, Martin-Luther-Universität Halle-Wittenberg, Von-Danckelmann-Platz 3, 06120 Halle, Germany*
[c]*Bergische Universität Wuppertal, FB C – Anorganische Chemie, Gaußstr. 20, D-42097 Wuppertal, Germany*
*Author to whom correspondence should be addressed. Electronic mail: yuyang5020@googlemail.com*

Abstract

In this work, relationships between the free volume and various fundamental physical properties (density, surface tension and transport properties) of ionic liquids were investigated. Two imidazolium ionic liquids 1-butyl-3-methylimidazolium tris(pentafluoroethyl)trifluoro phosphate ([$C_4$MIM][FAP]) and 1-butyl-3-methylimidazolium bis[bis(pentafluoroethyl)phosphinyl]imide ([$C_4$MIM][FPI]) were measured by positron annihilation lifetime spectroscopy (PALS). Changes of the ortho-positronium lifetime (*o*-Ps) with different states (amorphous and crystalline) were depicted as completely as possible. The mean local free (hole) volume $<v_h>$ was calculated from the *o*-Ps lifetime in amorphous state for the samples. Comparison between $<v_h>$ and specific volume obtained from the temperature dependent mass density gave the specific hole densities $N_f$ and the occupied volumes $V_{occ}$. Thermal expansion of hole volume was compared with molecular volume $V_M$ of [$C_4$MIM][FAP] and [$C_4$MIM][FPI] as well as five other ionic liquids from our previous works, a monotonically increasing correlation between the two quantities was displayed. Hole volume of [$C_4$MIM][FAP] sample from PALS experiment was compared with the result from surface tension according to Fürth hole theory, good agreement exhibited. The free volume obtained from this work was applied to Cohen-Turnbull fitting of viscosity for [$C_4$MIM][FPI] sample. The influence of the free volume to transport properties was investigated by the comparison of $<v_h>/V_M$ with the viscosity and conductivity for various ionic liquids. Correlation between the free volume and the molecular volume of ionic liquids were explained by a schematic free volume model.

Highlights

- Free volume thermal expansion correlates to the molecular volume for ionic liquids.
- A schematic free volume model is proposed to explain the volume correlation.
- The ratio, mean hole volume to molecular volume, correlates to transport properties.

## 1. Introduction

Ionic liquids are organic salts with low melting temperature (below 100 °C or even around room temperature) and ultra-low vapor pressure[1]. Many applications of this material rely on their excellent physical properties[1-6]. Up to $10^{18}$ possible candidates for this kind of material[7] requires microscopic molecular understanding of their structure and properties.

In our previous work [8, 9], the free volume in seven imidazolium ionic liquids was characterized and analyzed. The free volume $V_f$ representing the total interspace of ionic liquids relates to many fundamental physical properties of ionic liquids as presented below.

The mass density $\rho$ for one mole substance is:

$$\rho = \frac{M_A}{V_M \times N_A + V_{fm}} \qquad (1)$$

here, $M_A$ is the molar weight, $V_M$ is molecular volume, $N_A$ is Avogadro constant and $V_{fm}$ is molar free volume. When consider the $V_M$ as van der Waals volume $V_w$, $V_{fm}$ represents the total free space including interstitial volume. When $V_M$ is defined as the molecular volume in the crystalline structure $V_c$[10-12] (including the interstitial volume of the crystal structure), then the $V_{fm}$ is excess free volume that connects to transport properties[9]. In this work, $V_M$ takes the second meaning. For a certain substance, $M_A$, $V_M$ and $N_A$ are constants, so the mass density under fixed pressure and temperature is only a function of the free volume $V_{fm}$, which changes with pressure and temperature.

The surface tension of liquids is directly connected with the local free (hole) volume $v_h$ via the Fürth theory [13]. As shown in our previous study[8, 9], for the imidazolium ionic liquids, the mean hole volume $<v_h>$ from the positron annihilation lifetime spectroscopy (PALS) experiment is in accordance with the value calculated from surface tension by the method of Fürth theory. Large deviation comes from two ionic liquids [C$_4$MIM][Cl] and [C$_4$MIM][BF$_4$], possible reasons were discussed[9]. Considering the simple model used in the Fürth theory and the experimental error, the agreement is well.

It is well proved that the free volume is a main factor to influence the transport properties of liquids[14-18]. The Cohen-Turnbull equation[18] connects free volume with viscosity and conductivity. This equation is also valid for the ionic liquids[8, 9].

From the above discussion, it is clear that the free volume is important for the analysis of many fundamental physical properties of liquids. Then comes the question, of how to determine it. From the previous work for imidazolium ionic liquids, a correlation between the molecular volume $V_M$ and the hole volume was observed[9, 19]. Detailed comparison and quantitative correlation between the molecular volume $V_M$ and the mean hole volume were investigated [20]. Via the correlation between $V_f$ and $V_M$, the various physical properties of ionic liquids that connect to $V_M$ [21-23] can be interpreted under the term of free volume $V_f$.

This work is devoted to a comprehensive understanding of the free volume in amorphous ILs. The free volume was characterized as completely as possible for the two imidazolium ILs [C₄MIM][FAP] and [C₄MIM][FPI]. Calculated mean hole volume $<v_h>$ is applied for comparison with specific volume, the specific cavity density and occupied volume were obtained. Hole volume comparisons between PALS results and Fürth theory was made for [C₄MIM][FAP]. The Cohen-Turnbull model was applied to analyze the free volume dependence of viscosity for [C₄MIM][FPI]. Five other ionic liquids from previous work[8, 9] were also included here to investigate the molecular volume influence on free volume thermal expansion and transport properties.

2. Experimental

The two imidazolium ionic liquids in this work are [C₄MIM][FAP] (1-Butyl-3-methylimidazolium tris(pentafluoroethyl)trifluoro phosphat) and [C₄MIM][FPI] (1-Butyl-3-methylimidazolium bis[bis(pentafluoroethyl)phosphinyl]imide) as shown in Fig. 1. The water content was below 100 ppm for both ILs. The molar weight $M_A$ are 584.23 g/mol and 723.22 g/mol. Five other ionic liquids from the previous work[8, 9] are included for comparison,: 1-Butyl-3-methylimidazolium chloride ([C₄MIM][Cl]), 1-Butyl-3-methylimidazolium tetrafluoroborate ([C₄MIM][BF₄]), 1-Butyl-3-methylimidazolium hexafluorophosphate ([C₄MIM][PF₆]),1-Butyl-3-methylimidazolium bis(trifluoromethanesulfonyl)imide ([C₄MIM][NTf₂]) and 1-methyl-3-propylimidazolium bis(trifluoromethylsulfonyl)imide ([C₃MIM][NTf₂]).

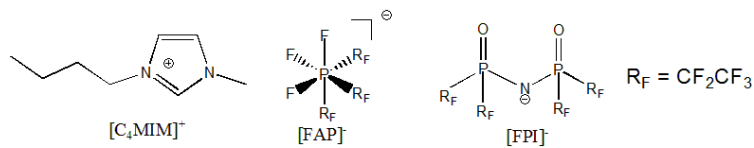

**Fig. 1** Chemical structure of [C₄MIM][FAP] and [C₄MIM][FPI].

Table 1 Ionic liquids characterization.

| Chemical Name | Source | Minimum Mass Fraction Purity | Purification Method | H₂O content (ppm) | Cl⁻ content (ppm) | Analysis Method |
|---|---|---|---|---|---|---|
| 1-Butyl-3-methylimidazolium tris(pentafluoroethyl)trifluoro phosphate | Merck KGaA | 0.99 | Dried at 70 °C for 12 – 48 h under stirring in vacuum | ⩽100 | ⩽50 | NMR |
| 1-Butyl-3-methylimidazolium bis[bis(pentafluoroethyl)phosphinyl]imide | | | | | | |

Due to the negligible vapor-pressure of ionic liquids, the samples can be measured under vacuum conditions in the positron annihilation lifetime spectroscopy (PALS) experiment. The radioactive source $^{22}Na$ was used to supply the positrons. This source was protected by 7 μm

Kapton foil before putting into a holder filled with liquid samples. The holder was put into a chamber with a pressure of $10^{-3}$ Pa. A fast-fast coincidence system with 310 ps time resolution and 25.7 ps analyzer channel width was applied for the measurement. The temperature range from 140 K to 315 K for [C$_4$MIM][FAP] and from 165 K to 315 K for [C$_4$MIM][FPI] with the step of 10 K was applied to obtain all the possible phase states (amorphous glass, supercooled liquid, normal liquid and crystalline phase) of the samples. Every experimental data point in the PALS experiment took 6 hours to collect at least $4*10^6$ counts for each spectrum. So the cooling/heating rate can be considered infinitesimal. Fast cooling from above the melting point $T_m$ to below the glass transition point $T_g$ was needed to obtain the supercooled state. This cooling rate was 5 K/min.

### 3. Results and Discussion

Table 2: The definition for individual volume.

| Volume | Definition | Unit |
|---|---|---|
| $V_s$ | specific volume | cm$^3$/g |
| $V_f$ | specific free volume | cm$^3$/g |
| $V_{fm}$ | molar free volume | cm$^3$/mol |
| $V_w$ | van der waals volume | Å$^3$ |
| $V_{occ}$ | occupied volume | Å$^3$ |
| $V_M$ | molecular volume | Å$^3$ |
| $V_c$ | crystalline volume | Å$^3$ |
| $v_f$ | free volume averaged to each molecule | Å$^3$ |
| $v_h$ | local free (hole) volume | Å$^3$ |

The relationship between respective volumes for one molecule is schematically shown in Fig. 2. $T_g$, $T_c$ and $T_m$ are the temperatures for glass transition, crystallization and melting, respectively. The van der Waals volume $V_w$ is the space occupied by a molecule, which is impenetrable to other molecules with normal thermal energies[24-26]. In this work, the molecular volume $V_M$ is defined to be equal to the crystalline volume $V_c$[10-12]. $V_I$ is defined as the interstitial volume in the crystal. The occupied volume is the volume extrapolated to 0 K for the glassy state, and the slight difference between $V_c$ and $V_{occ}$ comes from irregular dense packing[27]. The $V_c$ changes with temperature with much lower expansion coefficient than the free volume, so in this work, it is assumed constant. For the polymers, there is a correlation between $V_c$ and $V_w$: $V_c \sim 1.435 V_w$[28]. For the ionic liquids, $V_c$ (=$V_M$) ~ 1.410 $V_w$[29]. The value 1.435 for polymers and 1.410 for ionic liquids is close to $2^{1/2}$, the value corresponds to the minimum energy in Lennard-Jones (or 6-12) potential (at the position $r/\sigma=2^{1/6}$, then the volume ratio is $(2^{1/6})^3=1.414$) [25, 29, 30]. The $v_f$ is the free space averaged to each molecule, it is different from the local free (hole) volume $v_h$, which is the cavity in the real structure that can be seen from the PALS experiment. The hole volume $v_h$ is generated from coalescence of $v_f$ with statistics possibility[8, 9, 13]. The $V_f$, $V_{fm}$ are summation volumes of $v_f$ (or $v_h$) for one gram or one mole substance macroscopically.

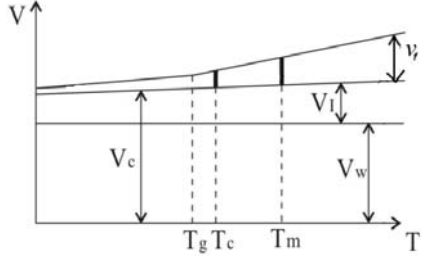

**Fig. 2** Schematic relationship of respective volumes of one molecule of ionic liquids and their temperature dependence.

### 3.1. Positron annihilation lifetime spectroscopy (PALS)

The positron will suffer thermalization and diffusion procedure after entering the material. Then positrons go through three decay channel with three distinct lifetimes. It would annihilate with an electron directly (with a lifetime denoted as $\tau_2$) or be trapped by a cavity forming a *para*-positronium (*p*-Ps, with a lifetime denoted as $\tau_1$) or an *ortho*-positronium (*o*-Ps, with a lifetime denoted as $\tau_3$). The *o*-Ps lifetime $\tau_3$ relates to the hole size (radius $r_h$), via pick-off annihilation, and a quantitative connection can be represented by the Tao-Eldrup equation[31, 32].

$$\tau_3 = \frac{0.5\,ns}{[1 - \frac{r_h}{r_h + \delta r} + \frac{1}{2\pi}\sin(\frac{2\pi r_h}{r_h + \delta r})]} \quad (2)$$

The parameter $\delta r = 1.66$ Å was determined empirically[32, 33].

Since the hole size distribution follows a thermal statistical law, the $\tau_3$ (=$1/\lambda_3$, $\lambda$ is annihilation rate) behaves as a distribution. Mean $<\tau_3>$ and its deviation $\sigma_3$ can be obtained from the software LifeTime 9.0[34] by decomposing the original spectrum $s(t)$ of the PALS measurement:

$$s(t) = I_1 \lambda_1 \exp(-\lambda_1 t) + \sum_{i=2,3} I_i \int_0^\infty \alpha_i(\lambda)\lambda \exp(-\lambda t) d\lambda \quad (3a)$$

with $\sum_{i=1,2,3} I_i = 1$, $\alpha_i(\lambda)\lambda d\lambda = \frac{1}{\sigma_i^*(2\pi)^{1/2}} \exp[-\frac{(\ln \lambda/\lambda_{i0})^2}{2\sigma_i^{*2}}] d\lambda$ (3b)

The software LifeTime9.0 utilizes non-linear least-squares fit of this function to the spectra, yielding the following annihilation parameters: $\tau_1 = 1/\lambda_1$, mean lifetime $<\tau_i>$, the mean dispersion (standard deviation) $\sigma_i$ of the corresponding lifetime distribution $\alpha_i(\tau) = \alpha_i(\lambda)d\lambda/d\tau = \alpha_i(\lambda)\lambda_2$, $<\tau_i> = \exp(\sigma^{*2}/2)/\lambda_{i0}$, and $\sigma_i = \sigma_i(\tau) = \sigma_i[\exp(\sigma_i^{*2})-1]^{0.5}$, and the relative intensities $I_i$.

The lifetime for normal amorphous materials is less than 4 ns, which is negligible comparing the structural relaxation time, e.g. 100 s at the glass transition temperature. So the positron lifetime takes a "snapshot" of the hole structure of the materials.

The radius distribution is:

$$n(r_h) = -\alpha_3(\lambda) d\lambda/dr_h = \frac{2\delta r}{(r_h + \delta r)^2}[1 - \cos(\frac{2\pi r_h}{r_h + \delta r})]\alpha_3(\lambda) \quad (4)$$

Then the hole volume distribution $g(v_h) = n(r_h)/4\pi r_h^2$. The number fraction of hole volumes between $v_h$ and $v_h + dv_h$ is $g_n(v_h) = g(v_h)/v_h$. So the number-weighted mean hole volume, $<v_h>$ and mean dispersion (standard deviation) $\sigma_h$, which characterized the $g_n(v_h)$ distribution corresponded to the first and second moment of the distribution respectively: $<v_h> = \int v_h g_n(v_h)dv_h/ \int g_n(v_h)dv_h$, $\sigma_h^2 = \int [v_h - <v_h>]^2 g_n(v_h)dv_h/ \int g_n(v_h)dv_h$.

The temperature dependent $o$-Ps lifetime parameters as well as the calculated hole volumes are shown for the two ionic liquids in this work. The data table can be found in the supporting information.

### 3.1.1. 1-Butyl-3-methylimidazolium tris(pentafluoroethyl)trifluoro phosphat [C$_4$MIM][FAP]

Fig. 3 shows the $o$-Ps lifetime variation with temperature for [C$_4$MIM][FAP] sample. The sample was first cooled (Cooling 1) from 300 K to 140 K, the temperature steps were 10 K. After a plateau of about 3.9 ns of $<\tau_3>$, a linear decrease with temperature was followed to 3.2 ns at 240 K. Then $<\tau_3>$ drops down to 1.5 ns and stays at that value. This is identified as crystallization[9]. The first heating procedure (Heating 1) was immediate after the Cooling 1, from 145 K to 315 K, in 10 K step. Before 265 K, the Heating 1 overlaps Cooling 1, meaning that the sample remains in the crystalline structure. At 275 K, the lifetime jumps from 1.5 ns to 3.9 ns, and the sample melts to the normal liquid state. It remains 3.9 ns afterwards. A fast cooling was applied from 330 K to 150 K with a rate of 5 K/min, this fast cooling resulted in a supercooled glassy phase, and the sample remains amorphous at 150 K. Then the second heating (Heating 2) from 150 K to 290 K in 10 K step was applied. The lifetime linearly increases with temperature from 1.74 ns to 1.96 ns, and a slope change at around 190 K, which is identified as the glass transition. At 210 K, a sudden jump to the crystalline lifetime happens, and the Heating 2 procedure duplicates the Heating 1 track. The $\sigma_3$ represents the deviation of $<\tau_3>$. The $T_k$ point indicates a leveling-off after melting in the normal liquid phase and is denoted as "knee temperature", since near this and higher temperatures, the structural relaxation time of the ionic liquids enters nanosecond region as the positron lifetime[9]. Due to a smearing of the hole wall, the positron lifetime can no more reflect correctly the hole size.

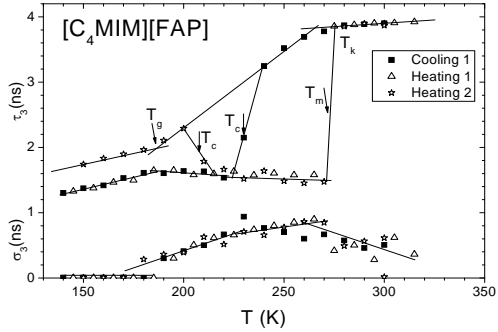

**Fig. 3** The mean $<\tau_3>$ and the standard deviation $\sigma_3$ of the o-Ps lifetime as a function of temperature $T$ during cooling (solid squares) and heating (empty triangles and stars) for [$C_4$MIM][FAP].

### 3.1.2. 1-Butyl-3-methylimidazolium bis[bis(pentafluoroethyl)phosphinyl]imide [$C_4$MIM][FPI]

Fig. 4 displays the PALS results of the [$C_4$MIM][FPI] sample. During cooling from 300 K to 170 K in steps of 10 K, the sample supercools until 220 K. A sudden drop from 3.2 ns to 1.9 ns indicates crystallization. After fast cooling in a rate of 5 K/min from 300 K to 165 K, the sample supercools at low temperature, showing an amorphous glassy state. In the heating, the lifetime linearly increased with temperature from 2 ns to 2.4 ns until a sudden fall at 215 K, indicating the transformation to the crystalline state, the lifetime remains around 1.6 ns until melting at 285 K and overlaps cooling procedure afterwards. The leveling-off at 270 K indicates the "knee temperature" $T_k$ in the cooling procedure.

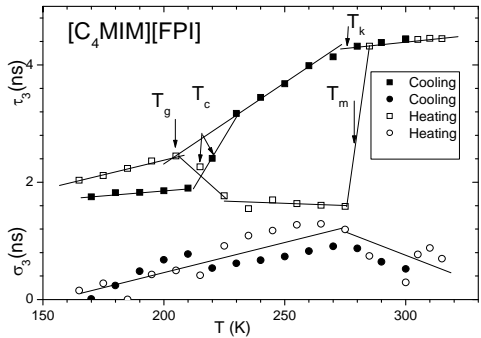

**Fig. 4** The mean $<\tau_3>$ and the standard deviation $\sigma_3$ of the o-Ps lifetime as a function of temperature $T$ during cooling (solid points) and heating (empty points) of [$C_4$MIM][FPI].

### 3.2. Local free (hole) volume from the PALS results:

Fig. 5 shows the hole volume parameters (mean hole volume $<v_h>$ and its deviation $\sigma_h$) calculated from o-Ps lifetime for the two samples. Only the hole volume in the amorphous phase other than the crystalline structure can be obtained. The mean hole volume $<v_h>$ linearly increases with temperature with a slope change at $T_g$ and levels off at $T_k$. The $<v_h>$ above $T_k$ does not mirror the real cavity structure anymore because the structural relaxation time of the substance reaches the region of the nanosecond [8, 9, 35, 36]. From the comparison with the pressure-volume-temperature (PVT) experiment[8, 9], the local free (hole) volume remains its linear thermal expansion above $T_k$ although it cannot be represented from the PALS experiment.

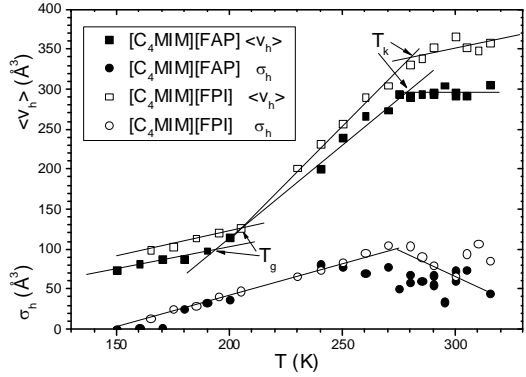

**Fig. 5** The mean $<v_h>$ and the standard deviation $\sigma_h$ of the hole size calculated from positron lifetime.

The mean hole volumes $<v_h>$ in the temperature range between $T_g$ to $T_k$ are compared with the specific volume $V_s$ for both samples as shown in Fig. 6. The specific volume $V_s$ is calculated from the reciprocal of the mass density: $V_s = 1/\rho$. The mass density was taken from the references for [C$_4$MIM][FAP] [37] and [C$_4$MIM][FPI] [38]. A linear fit of the form:

$$V_s = N_{fs} \times <v_h> + V_{occ} \qquad (5)$$

yields the parameters: for [C$_4$MIM][FAP], specific occupied volume $V_{occ}$=0.5479 cm$^3$/g, specific hole density $N_{fs}$ =0.198*10$^{21}$/g; for [C$_4$MIM][FPI], $V_{occ}$=0.5600cm$^3$/g, $N_{fs}$ =0.177*10$^{21}$/g. The linear behavior in equation (5) was confirmed in polymers [39-42] and low molecular weight glass formers[35, 36, 43]. Define the specific number density of molecule: $N_M=N_A/M_A$ ($N_A$: Avogadro constant; $M_A$(g mol$^{-1}$): molecular weight), then $N_{fs}/N_A$ for both samples are: 0.192 and 0.212. Then the ratio of hole number density to ion (cation and anion) density $N_{fs}/N_{ion}$ for both samples are: 0.096 for [C$_4$MIM][FAP], and 0.106 for [C$_4$MIM][FPI]. This ratio is close to the ratio from Fürth's work 1/11 (0.09) as the other imidazolium ionic liquids[9, 13].

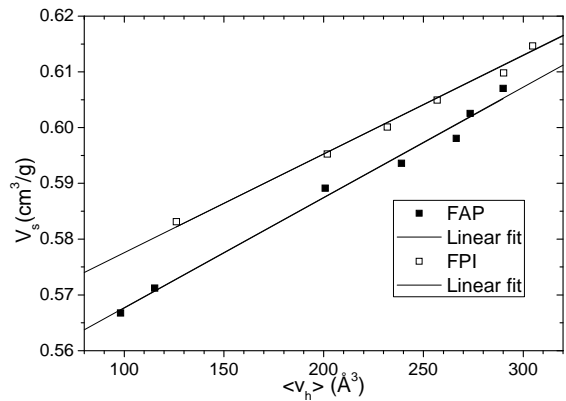

**Fig.6** Plot of the specific volume from mass density[37, 38] *vs.* the mean hole volume in the supercooled liquid state between $T_g$ and $T_k$ for [C$_4$MIM][FAP] and [C$_4$MIM][FPI]. The lines are linear fits of the data.

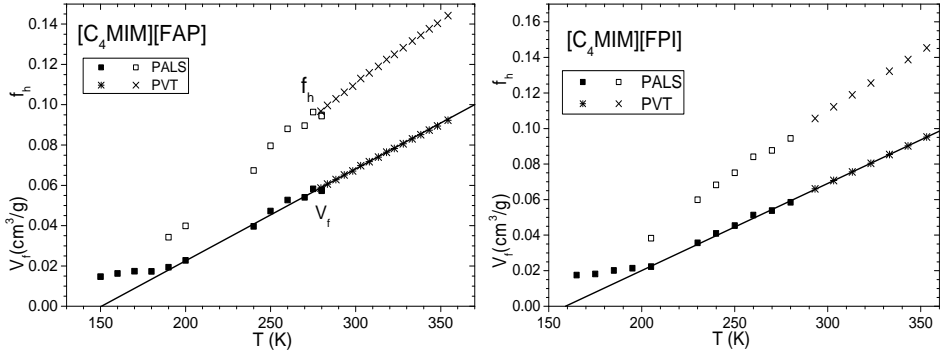

**Fig.7** Temperature dependence of the specific free volume $V_f$ and the free volume fraction $f_h$ of ionic liquids [C$_4$MIM][FAP] and [C$_4$MIM][FPI].

Fig 7 shows the temperature dependence of the free volume $V_f$ and the free volume fraction $f_h$ for [C$_4$MIM][FAP] and [C$_4$MIM][FPI]. Here, $V_f$ for PALS was calculated as the product of the mean hole volume and the specific hole density $<v_h> * N_{fs}$. The $V_f$ for PVT was calculated as subtraction of the specific volume from the occupied volume $V_s - V_{occ}$. The free volume fraction equals $V_f/V_s$. When the vacancies in the lattice-hole liquid model was considered as the Schottky-point defects, the hole fraction (as the equilibrium concentration) can be expressed as[44]:

$$f_h = A \exp(-\frac{H_h}{k_B T}) \qquad (6)$$

In equation (6): $A$ is a constant, ~~and~~ $H_h$ is the hole formation enthalpy, $T$ is temperature and $k_B$ is Boltzmann constant. The fitting of Fig. 7 delivers the result: for [C$_4$MIM][FAP], $A$=0.69 and $H_h$ = 4.57 kJ/mol; for [C$_4$MIM][FPI], $A$ = 0.77 and $H_h$ = 4.88 kJ/mol. The values are in agreement with the imidazolium ionic liquids in our previous work [9], but lower than the polymers [36, 43].

As shown in Fig 5, the mean hole volume $<v_h>$ exhibits a linear correlation with temperature (between $T_g$ and $T_k$ in the supercooled liquid phase) that can be represented as: $<v_h> = E_f T + C$. $E_f$ denotes the thermal expansion of the mean hole volume. In Fig 8, the $E_f$ was compared with $V_M$ for [C$_4$MIM][FAP] and [C$_4$MIM][FPI] as well as five other ionic liquids from our previous works: [C$_4$MIM][Cl], [C$_4$MIM][BF$_4$], [C$_4$MIM][PF$_6$], [C$_4$MIM][NTf$_2$] and [C$_3$MIM][NTf$_2$] [8, 9]. The value of $V_M$ was chosen as the value of $V_{m,scaled}$ from the Beichel's work [20]. Obvious correlation between $E_f$ and $V_M$ can be discerned. This correlation will be explained by a schematic free volume model in section 3.5 of this work.

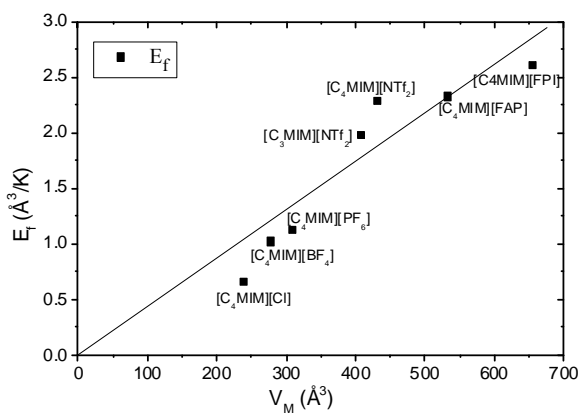

**Fig.8** The comparison between mean hole volume thermal expansion $E_f$ and molecular volume $V_M$ [20] for seven imidazolium ionic liquids. The solid line is a guide to the eye.

### 3.3. Hole volume comparison with Fürth theory for the [C$_4$MIM][FAP] sample:

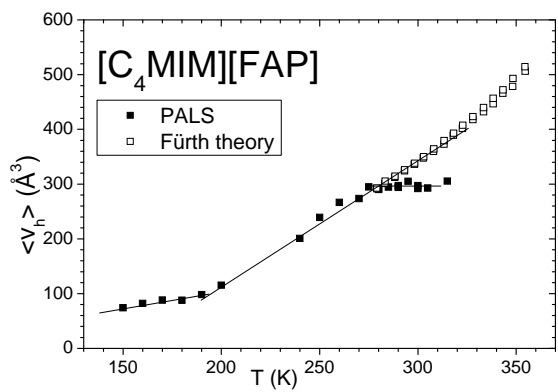

**Fig. 9** Comparison of the mean hole volume $<v_h>$ of [C$_4$MIM][FAP] from PALS (solid symbols) and the cavity volume as calculated by Fürth theory [8, 9, 13] (empty symbols) from surface tension data [37].

The Fürth theory is based on a simple idea that the formation of a hole of spherical shape is equal to the sum of work to be done against the surface tension and the pressure. The single hole will coalesce into complex larger cavity with the maximum possibility of containing 11 single holes[8, 9, 13]. Fig. 9 compared the local free (hole) volume from PALS experiment and the calculated value from surface tension[37] by Fürth theory[8, 9, 13] for the [C$_4$MIM][FAP] sample. The good agreement was as earlier shown for other imidazolium ionic liquids[8, 9].

### 3.4. Viscosity for the [C$_4$MIM][FPI] sample and free volume influence of transport properties of ionic liquids

The free volume and the diffusion coefficient of molecules can be connected by the Cohen-Turnbull equation[18, 45, 46]: $D \propto T^{1/2}\exp(-\gamma V_f^*/V_f)$. $V_f^*$ is the minimum required volume of the void, $\gamma$ is a constant for a single substance and is an overlap factor which should lie between 0.5 and unity, $T$ is the temperature, and $V_f$ is the specific free volume. From the Stokes-Einstein equation $D \propto k_B T/\eta$, here $k_B$ is the Boltzmann constant, the viscosity $\eta$ can be related to the free volume:

$$\eta = CT^{1/2} \exp(\gamma V_f^*/V_f) \quad (7)$$

with $C$ as a constant.

The Vogel-Fulcher-Tammann (VFT) [47-49] equation of viscosity can be obtained from equation (7) when assuming that the free volume expands linearly with temperature, $V_f = E_f(T-T_0')$:

$$\eta = \eta_0 T^{1/2} \exp[B/(T-T_0)] \quad (8)$$

Comparing both equations: $T_0=T_0'$, $\eta_0=C$ and $B=\gamma V_f^*/E_f$.

Fig. 10 shows the VFT and CT fitting of viscosity for the [C$_4$MIM][FPI], viscosity data were obtained from reference[38]. The value for the CT fitting parameter of equation (7) are: Ln($C$)= -12.5, $\gamma V^*$=0.513 cm$^3$g$^{-1}$. For the VFT equation (8), the fitting result are: Ln($\eta_0$)= -12.4, $B$=1017 K, $T_0$=160.4 K. The coefficient of determination of $R^2$=0.99999 for both fits.

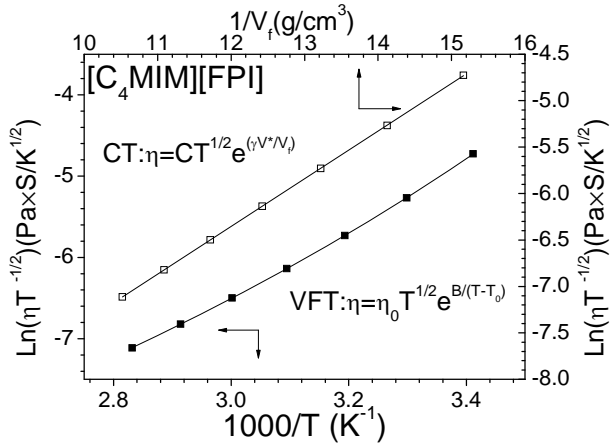

**Fig. 10** VFT (Vogel-Fulcher-Tammann, solid squares) and CT (Cohen-Turnbull, empty squares) fitting of the viscosity for [C$_4$MIM][FPI].

A more direct and intuitive free volume influence to transport properties is manifested by comparison of the ratio $<v_h>/V_M$ with viscosity and conductivity, as shown in Fig 11 for different ionic liquids at two chosen temperature 270 K and 320 K. The mean hole volume $<v_h>$ used here is from interpolation or extrapolation at 270 K and 320 K of the linear fitting of $<v_h>$ versus $T$ at the supercooled liquid state for the 6 samples: [C$_4$MIM][Cl], [C$_4$MIM][BF$_4$], [C$_4$MIM][PF$_6$], [C$_3$MIM][NTf$_2$], [C$_4$MIM][NTf$_2$], and [C$_4$MIM][FPI] [8, 9]. The viscosity and conductivity data are from the interpolation or extrapolation at 270 K and 320 K of the VFT fitting of the experimental data[8, 9, 38, 50-55] of the ionic liquids. From Fig. 11, the viscosity decreases with increased $<v_h>/V_M$ while the conductivity increases with $<v_h>/V_M$. It is in agreement with the hole theory that diffusion happens when a hole large enough opens for the molecules with volume $V_M$. [18]

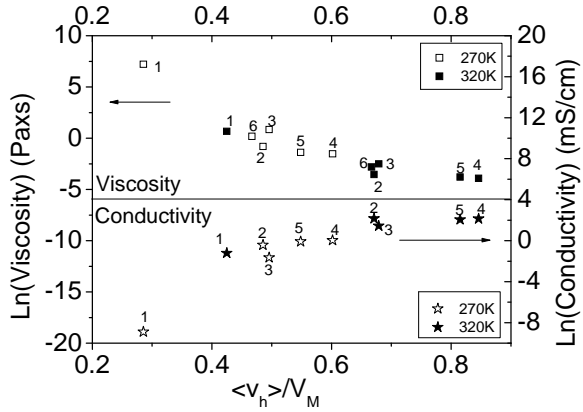

**Fig. 11** The comparison of the ratio $<v_h>/V_M$ with viscosity (above, square points) and conductivity (below, star points) for different ionic liquids (numbered in the graph) at two chosen temperature 270 K (empty points) and 320 K (solid points). The number denotes the ionic liquids: 1. [C$_4$MIM][Cl], 2. [C$_4$MIM][BF$_4$], 3. [C$_4$MIM][PF$_6$], 4. [C$_3$MIM][NTf$_2$], 5. [C$_4$MIM][NTf$_2$], and 6. [C$_4$MIM][FPI].

### 3.5. Free volume model

The chemical structure (Fig.1) indicates a larger molecular volume of [C$_4$MIM][FPI] than [C$_4$MIM][FAP]. The Fig. 5 shows a larger $<v_h>$ of [C$_4$MIM][FPI], as well as a larger thermal expansion (slope of the $<v_h> \sim T$ line) of [C$_4$MIM][FPI] than [C$_4$MIM][FAP]. The $v_f$ can also be calculated (from extrapolation of Fig. 7) for the two samples at 300 K: 66 Å$^3$ and 83 Å$^3$ for [C$_4$MIM][FAP] and [C$_4$MIM][FPI] respectively. So the correlation between free volume and molecular volume [20] is also valid for the two ionic liquids in this work.

A schematic model to explain the correlation assumption between $v_f$ and $V_M$ is displayed in Fig. 12. The particles are simplified as spheres, two particles are separated by distance $d$, then the free volume for each particle $v_f$ is supposed to be the larger sphere subtracted the molecular volume: $4\pi/3(r_V^3 - r_M^3)$, here $r_V = r_M + d/2$. So $v_f = \pi d(2r_M^2 + r_M d + d^2/6)$. When $d \ll r_M$, $v_f \cong 2\pi d r_M^2 \sim dV_M^{2/3}$. The free volume $v_f$ has relationship with $V_M$ and $d$, here the distance $d$ under certain temperature depends on the force and molecular movement between particles. Because the $v_f$ linearly increase with temperature, so $d$ linearly changes with $T$, that is $d = CT$, $C$ is constant, so $v_f \sim CV_M^{2/3}T$, then the correlation between free volume thermal expansion and molecular volume $V_M$ (Fig. 8) can be obtained (because the hole number density is constant, the thermal expansion for $<v_h>$ and $v_f$ is the same). Here the sphere is considered for simplification, when real particles are introduced, molecular structure and configuration should be accounted in.

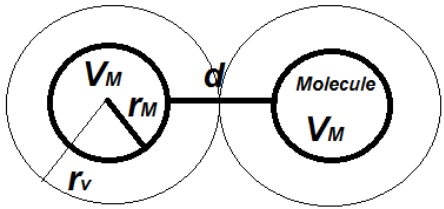

**Fig. 12** Schematic model for the $v_f$ and $V_M$.

## 4. Conclusion

The local free volume was studied as completely as possible for the two imidazolium ionic liquids [C$_4$MIM][FAP] and [C$_4$MIM][FPI]. In different phases, the *o*-Ps lifetime parameters show different behaviors: the distinct low mean lifetime $<\tau_3>$ indicates the crystalline phase; slope change in amorphous phase signifies a glass transition and a leveling off at $T_k$ means fast structural relaxation and the positron lifetime does not represent the hole size correctly anymore.

Good agreement between the Fürth theory and the PALS experimental results was obtained. The influence of the free volume to transport properties was investigated by the Cohen-Turnbull fitting of viscosity for the ionic liquid [C$_4$MIM][FPI], a more direct and intuitive free volume influence to transport properties is manifested by comparison of the ratio $<v_h>/V_M$ with viscosity and conductivity for different ionic liquids under different temperature.

A schematic model was applied to explain the correlation between $v_f$ and $V_M$ as well as the relationship between free volume thermal expansion and $V_M$. One thing to be emphasized is: $v_f$ is not the cavity volume in real structure but the average free volume for each molecule. In the real structure, due to the mobility of the molecules, the $v_f$ coalesces following a thermal dynamic law and forms complex larger holes.

The free volume is not only the individual hole that affects the transport property, but also space that connects potential energy between particles.

From the discussion above, the free volume shows certain regularities and affects transport properties for the ionic liquids. To verify their nature, the hole volumes of various series of ionic liquids other than with imidazolium cations are needed in the future work.

*Acknowledgements*

*The authors would like to thank Prof. Dr Jerzy Kansy (Institute of Materials Science, Silesian University, Katowice) for supplying the analyzing routine LifeTime 9.0. Dr. N. Ignat'ev (Merck KGaA) and Merck KGaA (Darmstadt, Germany) are acknowledged for providing the [C$_4$MIM][FAP] and [C$_4$MIM][FPI] samples. Dr. Yang Yu acknowledges the financial support from the National Natural Science Foundation of China (Grant No.: 11247220) and the Science Research Startup Foundation from the Nanjing University of Information Science & Technology (Grant No: S8112078001).*